\documentclass[aps,prd,twocolumn,superscriptaddress,nofootinbib]{revtex4-1}

\usepackage{latexsym}
\usepackage{amsmath}
\usepackage{amssymb}
\usepackage{amsfonts}
\usepackage{bm}
\usepackage{physics}

\usepackage{color}
\definecolor{purple}{rgb}{0.5,0,0.5}
\definecolor{blue}{rgb}{0.0,0,0.9}
\definecolor{prdblue}{rgb}{0.133,0.118,0.498}
\usepackage[colorlinks=true, pdfstartview=FitV, linkcolor=prdblue, citecolor= prdblue, urlcolor=prdblue]{hyperref}

\usepackage{supertabular}
\usepackage{placeins}
\usepackage{epsfig}
\usepackage{graphicx}

\usepackage{soul}

\usepackage{CJKutf8}

\begin{document}
\begin{CJK*}{UTF8}{gbsn}
%\preprint{}

\title{Parton distribution functions of ground state mesons composed of $c$ or $b$ quarks}


\author{Qian Wu (吴迁)}
\email[]{qwu@nju.edu.cn}
\affiliation{School of Physics, Nanjing University, Nanjing, Jiangsu 210093, China}

\author{Zhu-Fang Cui (崔著钫)}
\email[]{phycui@nju.edu.cn}
\affiliation{School of Physics, Nanjing University, Nanjing, Jiangsu 210093, China}

\author{Jorge Segovia}
\email[]{jsegovia@upo.es}
\affiliation{Dpartamento de Sistemas F\'isicos, Qu\'imicos y Naturales, Universidad Pablo de Olavide, E-41013 Sevilla, Spain}


\date{\today}

\begin{abstract}
The valence quark parton distribution functions (PDFs) of all ground state heavy mesons composed of $b$ or $c$ quarks, are discussed; namely, the pseudoscalar $\eta_c(1S)$, $\eta_b(1S)$ and $B_c$, together with the corresponding vector ones, $J/\psi$, $\Upsilon(1S)$ and $B_c^\ast$.
We use a QCD-inspired constituent quark model, which has been applied with success to conventional heavy mesons, so that one advantage here is that all parameters have already been fixed by previous studies.
The wave functions of the heavy mesons in the rest frame are obtained by solving the Schr\"odinger equation, then boosted to its light-front, based on Susskind's Lorentz transformation. The PDFs at the hadron scale, are then obtained by integrating out the transverse momenta of the modulus square of the light-front wave function.
Our study shows how the valence quark distributions differ between pseudoscalar and vector mesons, as well as among charmonia, bottomonia and bottom-charmed mesons. Comparisons with other theoretical calculations demonstrate that the PDFs obtained herein are in general narrower but align well with the expected patterns. Moreover, each PDF's point-wise behavior is squeezed with respect to the scale-free parton-like PDF.

\end{abstract}

%\pacs{}
%\keywords{}

\maketitle

%%%%%%%%%%%%%%%%%%%%%%%%%%%%%%%%%%%%%%%%%%%%%%%%%%%%%%%%%%%%%%%%%%%%%%%%%%%%%%%%%
%%%%%%%%%%%%%%%%%%%%%%%%%%%%%%%%%%%%%%%%%%%%%%%%%%%%%%%%%%%%%%%%%%%%%%%%%%%%%%%%%

\end{CJK*}

\section{INTRODUCTION}
\label{sec:Intro}
Mesons composed of heavy quarks, such as charm ($c$) and bottom ($b$), provide great frameworks for systematically studying characters of strong interaction, as well as testing quantum chromodynamics (QCD). It stems from their large masses compared to the inherent QCD scale, $\Lambda_{\rm QCD}$. Moreover, the velocities of the (anti-) heavy quarks in these systems are low enough to allow us to use some non-relativistic potential to describe their properties~\cite{QuarkoniumWorkingGroup:2004kpm,Brambilla:2004jw,Brambilla:2010cs,Rothkopf:2019ipj,Akamatsu:2020ypb,Shuryak:2021yif}. On the other hand, quarkonia, \emph{i.e.}, the bound states of a quark $Q$ and its anti-quark $\bar Q$, are very special. The energy scales involved span from the hard region, where perturbation methods can work, to the low energy region, that is dominated by confinement and complicated non-perturbative dynamics. This makes them unique in many topics, such as the nature of the confinement mechanism, testing the Standard Model (SM) of particle physics, explaining the emergence of exotics $X$, $Y$, $Z$ states, phase diagram of nuclear matter, and then cosmology as well as our universe, etc.~\cite{Gross:2022hyw}. For example, discovery of $J/\psi$, the vector ground state of charmonium ($c\bar c$), drastically changed and shaped SM: it represented not only the discovery of the $c$ quark, but also a confirmation of the quark model.

Parton distribution functions (PDFs) describe how the light-front momentum of a hadron is shared among its constituent partons, namely quarks and gluons. In other words, PDFs provide the probability of finding a particular parton inside a hadron, carrying a specific fraction of the total momentum, when the hadron is moving at a very high speed that is close to light. These functions are essential in understanding the internal structure of hadrons, play a key role in making predictions for high-energy collisions, and help to connect fundamental parton-level interactions, governed by QCD, with the observed hadron properties in experiments~\cite{Proceedings:2020fyd,Cui:2020tdf,Yin:2023dbw,Yu:2024ovn}. 

On the other hand, the precise determination of nucleon PDFs from experimental data~\cite{Gao:2017yyd, Kovarik:2019xvh, Ethier:2020way} requires analyzing the cross sections of inclusive $c$ and $b$ quarks production in deep inelastic scattering (DIS), and then the derived structure functions $F_2^{c\bar c}$ and $F_2^{b\bar b}$~\cite{H1:2015ubc, H1:2018flt}. The existence of a nonperturbative intrinsic heavy quark component in the nucleon is a rigorous prediction of QCD, thus $c$ and $b$ quarks provide a fundamental property of the wave functions of bound states. While the extrinsic contributions to the heavy quark PDFs  are most important at low $x$, the intrinsic ones take charge at high $x$, since the latter are kinematically dominated by the regime where the $|uudQ\bar Q\rangle$ state is minimally off shell, corresponding to equal rapidities of the constituent quarks~\cite{Brodsky:2015fna}. Recently, a phenomenological fit to a selection of available data may be interpreted as supporting the existence of a small intrinsic charm component in the proton~\cite{Ball:2022qks}. This has led to significant attention on the PDFs of heavy quarks within the nucleon. However, there is still limited experimental and theoretical knowledge regarding the PDFs of mesons with heavy valence quarks.

Experimentally, the situation is expected to improve with upcoming programs that focus on the production of heavy mesons in, for example, the high-luminosity Large Hadron Collider (HL-LHC)~\cite{Feng:2022inv} or the electron-ion colliders in the USA (EIC)~\cite{AbdulKhalek:2021gbh} and China (EicC)~\cite{Anderle:2021wcy}. On the other hand, in the last few years, there have been several theoretical works that discussed PDFs of heavy mesons. In Refs.~\cite{Albino:2022gzs, Almeida-Zamora:2023bqb}, the authors propose an algebraic model for the quark Dyson-Schwinger equation and for the pseudoscalar meson's Bethe-Salpeter equation to compute, among other quantities, the heavy quark PDF of $\eta_c(1S)$, $\eta_b(1S)$ and $B_c$ mesons. Ref.~\cite{Almeida-Zamora:2023rwg} attempts to compute the leading-twist general parton distributions of heavy vector mesons within the continuum Schwinger function method. Using a light-front base quantization approach and solving the effective Hamiltonian, Ref.~\cite{Lan:2019img} provides PDFs of various heavy mesons. Moreover, within a light-front quark model, some structure properties of heavy mesons are discussed in Ref.~\cite{Arifi:2024mff}.

Recently, we calculated pion PDF with a QCD-inspired constituent quark model (CQM)~\cite{Wu:2022iiu}. The idea is to calculate the light-front wave function (LFWF) of pion by boosting its CQM eigenfunction at the rest frame to the light-front, in which the DIS occurs, using an inverse Lorentz transformation. Then, the parton distribution of the valence $u$ quark in the pion at the hadron scale\footnote{At which all properties (eg., light-front momentum) of a given hadron are  carried by its valence quark~\cite{Yin:2023dbw}. In our case, it is $0.313\,\text{GeV}$.} is obtained and evolved by the so-called Dokshitzer-Gribov-Lipatov-Altarelli-Parisi (DGLAP) evolution equations to the relevant energy scales, at which one can compare with experimental data. The obtained results about the pion quark distribution at $Q^2 = 20\,\text{GeV}^2$ and the $F_2^\pi$ structure function are in fair agreement with the experimental measurements once the mass of the dressed valence light quark is calibrated accordingly. This deficiency can be traced back to the limitations of the CQM in describing the fine details of light meson properties. However, the aforementioned weakness of this approach in principle should not occur in the heavy meson sector. Thus, in this work, we generalize it to investigate the leading-twist PDFs of pseudoscalar and vector ground states of all mesons composed with $c$ or $b$ quarks.

The presentation is arranged as follows. In Sect.~\ref{sec:Theory}, we introduce CQM, the numerical method to solve the resulting Sch\"odinger equation, and how the wave functions calculated in the rest frame are transformed into infinite momentum or light-front frame. The PDFs of ground state pseudoscalar and vector heavy mesons, with hidden charm or bottom as well as the charmed-bottom case, are provided in Sect.~\ref{sec:Results}. Finally, we provide a summary and concluding remarks in Sect.~\ref{sec:Summary}.

%%%%%%%%%%%%%%%%%%%%%%%%%%%%%%%%%%%%%%%%%%%%%%%%%%%%%%%%%%%%%%%%%%%%%%%%%%%%%%%%%
%%%%%%%%%%%%%%%%%%%%%%%%%%%%%%%%%%%%%%%%%%%%%%%%%%%%%%%%%%%%%%%%%%%%%%%%%%%%%%%%%

\section{THEORETICAL FRAMEWORK}
\label{sec:Theory}

%%%%%%%%%%%%%%%%%%%%%%%%%%%%%%%%%%%%%%%%%%%%%%%%%%%%%%%%%%%%%%%%%%%%%%%%%%%%%%%%%

\subsection{Constituent Quark Model}
\label{subsec:CQM}
It is well known that the quark model has been very successful in explaining the properties of hadrons since its introduction in the 1960s by Gell-Mann and Zweig~\cite{GellMann:1964nj, Zweig:1964CERN}. For instance, focusing on the meson sector, the study of quarkonia within this theoretical framework found that heavy quark systems are properly described by non-relativistic potential quark models reflecting the dynamics expected from QCD~\cite{Eichten:1978tg, Eichten:1979ms}. Furthermore, the \emph{a priori} complicated light meson sector was surprisingly well reproduced in its bulk properties by means of a universal perturbative one-gluon exchange (OGE) plus a linear confining potential~\cite{Godfrey:1985xj}. However, the dynamics of the light quark sector is expected to be dominated by the non-perturbative spontaneous breaking of chiral symmetry, and consequently constituent light quarks should interact through the exchange of Goldstone bosons~\cite{Glozman:1997fs, Segovia:2008zza, Fernandez:2021zjq}. Therefore, for the light quark sector, hadrons may be described as systems of confined quarks (anti-quarks) interacting through gluons and boson exchanges, whereas heavy hadrons are systems of confined quarks interacting through gluon exchanges only.

The CQM we use here is proposed in~\cite{Vijande:2004he} and extensively reviewed in~\cite{Valcarce:2005em, Segovia:2013wma, Ortega:2012rs}.
It has been applied with success to conventional mesons containing heavy quarks, describing a wide range of physical observables that concern spectra~\cite{Segovia:2015dia, Ortega:2020uvc}, strong decays~\cite{Segovia:2009zz, Segovia:2013kg}, hadronic transitions~\cite{Segovia:2014mca, Martin-Gonzalez:2022qwd} as well as electromagnetic and weak reactions~\cite{Segovia:2012yh}. The advantage of using such an approach with a relatively long history is that all model parameters are already well constrained by previous works, and consequently, from this perspective, one can provide relatively reliable, and in some sense parameter-free predictions, since in our calculations no fine tuning or new parameter is introduced. The OGE potential contains central, tensor and spin-orbit contributions given by~\cite{Segovia:2008zz}
\begin{subequations}
\begin{align}
V_\text{OGE}^\text{C}(\vec{r}_{ij}) &= \frac{1}{4}\alpha_{s}(\vec{\lambda}_{i}^{c}\cdot \vec{\lambda}_{j}^{c}) \Bigg[ \frac{1}{r_{ij}} \nonumber \\
&
-\frac{1}{6m_{i}m_{j}} (\vec{\sigma}_{i}\cdot\vec{\sigma}_{j})  \frac{e^{-r_{ij}/r_{0}}}{r_{ij}r_{0}^{2}} \Bigg], \\
V_\text{OGE}^\text{T}(\vec{r}_{ij}) &= -\frac{1}{16}\frac{\alpha_{s}}{m_{i}m_{j}} (\vec{\lambda}_{i}^{c}\cdot\vec{\lambda}_{j}^{c}) \Bigg[  \frac{1}{r_{ij}^{3}} \nonumber \\
&
-\frac{e^{-r_{ij}/r_{g}}}{r_{ij}}\left(  \frac{1}{r_{ij}^{2}}+\frac{1}{3r_{g}^{2}}+\frac{1}{r_{ij}r_{g}}\right) \Bigg] S_{ij}, \\
V_\text{OGE}^\text{SO}(\vec{r}_{ij}) &=  -\frac{1}{16}\frac{\alpha_{s}}{m_{i}^{2}m_{j}^{2}}(\vec{\lambda}_{i}^{c} \cdot \vec{\lambda}_{j}^{c}) \nonumber \\
&
\times \Bigg[\frac{1}{r_{ij}^{3}}-\frac{e^{-r_{ij}/r_{g}}} {r_{ij}^{3}} \left(1+\frac{r_{ij}}{r_{g}}\right)\Bigg] \nonumber \\
&
\times \Big[((m_{i}+m_{j})^{2}+2m_{i}m_{j})(\vec{S}_{+}\cdot\vec{L}) \nonumber \\
&
+ (m_{j}^{2}-m_{i}^{2}) (\vec{S}_{-}\cdot\vec{L}) \Big],
\label{eq:OGEpot}
\end{align}
\end{subequations}
with $S_{ij}=3(\vec{\sigma}_{i}\cdot\hat{r}_{ij})(\vec{\sigma}_{j}\cdot \hat{r}_{ij})-\vec{\sigma}_{i}\cdot\vec{\sigma}_{j}$ the quark tensor operator, whereas $r_0\equiv r_{0}(\mu_{ij})=\hat{r}_{0}\frac{\mu_{nn}}{\mu_{ij}}$ and $r_{g}\equiv r_{g}(\mu_{ij})=\hat{r}_{g}\frac{\mu_{nn}}{\mu_{ij}}$ are regulators which depend on $\mu_{ij}$, the reduced mass of the quark-antiquark pair, and being $\mu_{nn}=m_n/2=(313/2)\,\text{MeV}$. The contact term of the central potential has been regularized as
\begin{equation}
\delta(\vec{r}_{ij}) \approx \frac{1}{4\pi  r_{0}^{2}}\frac{e^{-r_{ij}/r_{0}}}{r_{ij}}.
\label{eq:delta}
\end{equation}

The wide energy range needed to provide a consistent description of light, strange and heavy mesons requires an effective scale-dependent strong coupling constant~\cite{Cui:2019dwv,Deur:2023dzc}; we use the one suggested in Ref.~\cite{Vijande:2004he}, whose expression is given by
\begin{equation}
\alpha_{s}(\mu_{ij})=\frac{\alpha_{0}}{\ln\left(
\frac{\mu_{ij}^{2}+\mu_{0}^{2}}{\Lambda_{0 }^{2}} \right)},
\end{equation}
in which $\alpha_{0}$, $\mu_{0}$ and $\Lambda_{0}$ are parameters of the model determined by a global fit to the meson spectra.

The different pieces of the confinement potential are
\begin{subequations}
\begin{align}
&
V_{\rm CON}^{\rm C}(\vec{r}_{ij})=\left[-a_{c}(1-e^{-\mu_{c}r_{ij}})+\Delta
\right] (\vec{\lambda}_{i}^{c}\cdot\vec{\lambda}_{j}^{c}), \\
&
V_{\rm CON}^{\rm SO}(\vec{r}_{ij}) =
-(\vec{\lambda}_{i}^{c}\cdot\vec{\lambda}_{j}^{c}) \frac{a_{c}\mu_{c}e^{-\mu_{c}
r_{ij}}}{4m_{i}^{2}m_{j}^{2}r_{ij}} \nonumber \\
&
\times
\left[((m_{i}^{2}+m_{j}^{2})(1-2a_{s}) + 4m_{i}m_{j}(1-a_{s}))(\vec{S}_{+}
\cdot\vec{L}) \right. \nonumber \\
&
\left. \quad\,\, +(m_{j}^{2}-m_{i}^{2}) (1-2a_{s}) (\vec{S}_{-}\cdot\vec{L})
\right],
\end{align}
\end{subequations}
where $a_{s}$ controls the mixture between scalar and vector Lorentz structures of confinement. At short distances, this potential presents a linear behavior with an effective confinement strength $\sigma = -a_{c} \, \mu_{c} \, (\vec{\lambda}^{c}_{i}\cdot \vec{\lambda}^{c}_{j})$, while it becomes constant at large distances with a threshold defined by $V_{\rm thr}=(-a_{c}+ \Delta)(\vec{\lambda}^{c}_{i}\cdot \vec{\lambda}^{c}_{j})$. There are no $q\bar{q}$ bound states for energies exceeding this threshold. At this point, the system transitions from a color string configuration between two static color sources to a pair of static mesons, resulting from the breaking of the color string and the preferred decay into hadrons.

In order to solve the Sch\"odinger equation and find the quark-antiquark bound states, we use the Gaussian Expansion Method (GEM)~\cite{Hiyama:2003cu} which provides enough precision and simplifies the evaluation of matrix elements. This procedure provides the radial wave function solution of the Schr\"odinger equation as an expansion in terms of Gaussian basis functions
\begin{equation}
R_{\alpha}(r)=\sum_{n=1}^{n_{max}} c_{n}^\alpha \phi^G_{nl}(r),
\end{equation}
where $\alpha$ refers to the channel's quantum numbers. The coefficients,
$c_{n}^\alpha$, and the eigenvalue, $E$, are determined from the Rayleigh-Ritz
variational principle
\begin{equation}
\sum_{n=1}^{n_{max}} \left[\left(T_{n'n}^\alpha-EN_{n'n}^\alpha\right)
c_{n}^\alpha+\sum_{\alpha'}
\ V_{n'n}^{\alpha\alpha'}c_{n}^{\alpha'}=0\right],
\end{equation}
where $T_{n'n}^\alpha$, $N_{n'n}^\alpha$ and $V_{n'n}^{\alpha\alpha'}$ are the
matrix elements of the kinetic energy, the normalization and the potential,
respectively. 

\begin{table}[!t]
\caption{\label{tab:CQM-parameters} The CQM's parameters fitted over all meson spectra and relevant to the heavy quark sectors.}
\begin{ruledtabular}
\begin{tabular}{llr}
Quark masses & $m_c$ (MeV) & $1763$ \\
             & $m_b$ (MeV) & $5110$ \\
\hline
OGE & $\hat{r}_0$ (fm) & $0.181$ \\
    & $\hat{r}_g$ (fm) & $0.259$ \\
    & $\alpha_{0}$ & $2.118$ \\
    & $\Lambda_{0}$ (fm$^{-1}$) & $0.113$ \\
    & $\mu_{0}$ (MeV) & $36.98$ \\
\hline
Confinement & $a_{c}$ (MeV) & $507.4$ \\
            & $\mu_{c}$ (fm$^{-1}$) & $0.576$ \\
            & $\Delta$ (MeV) & $184.432$ \\
            & $a_s$ & $0.81$ \\
\end{tabular}
\end{ruledtabular}
\end{table}

The model parameters relevant to this work are listed in Table~\ref{tab:CQM-parameters}. Details about how to fit them can be found in, for instance, Refs.~\cite{Vijande:2004he, Segovia:2008zz, Segovia:2013wma}. Table~\ref{tab:QQspectra} lists the masses of the heavy mesons whose valence quark PDF are calculated herein. For comparison, the experimental data reported in the Review of Particle Physics of Particle Data Group (PDG) are also listed~\cite{ParticleDataGroup:2024cfk}.

It is important to highlight that there are two types of theoretical uncertainties in our results: one is intrinsic to the numerical algorithm and the other is related to the way the model parameters are fixed. The numerical error is negligible and, as mentioned above, the model parameters are adjusted to reproduce a certain number of hadron observables within a determinate range of agreement with experiment. It is therefore difficult to estimate an error for these parameters and consequently for the quantities calculated using them.

\begin{table}[!t]
\caption{\label{tab:QQspectra} Masses, in MeV, of the heavy mesons whose valence quark PDF are calculated herein. The experimental values are taken from Ref.~\cite{ParticleDataGroup:2024cfk}.}
\begin{ruledtabular}
\begin{tabular}{ccc}
Meson & Theory & Experiment \\
\hline
$\eta_c(1S)$    & $2990$ & $2984.1\pm0.4$ \\
J$/\psi$    & $3096$ & $3096.900\pm0.006$ \\
$B_c$       & $6277$ & $6274.47\pm0.27$ \\
$B_c^*$     & $6328$ & $\cdots$ \\
$\eta_b(1S)$    & $9455$ & $9398.7\pm2.0$ \\
$\Upsilon(1S)$  & $9502$ & $9460.40\pm0.09$ \\
\end{tabular}
\end{ruledtabular}
\end{table}

%%%%%%%%%%%%%%%%%%%%%%%%%%%%%%%%%%%%%%%%%%%%%%%%%%%%%%%%%%%%%%%%%%%%%%%%%%%%%%%%%

\subsection{Meson wave function in the light-front}
\label{subsec:LighconeWF}

In DIS processes, a hadron can be regarded as moving with infinite momentum. Susskind proposed that the infinite momentum frame limiting procedure is essentially a change from the laboratory time and $z$ coordinates to the light-cone time and space coordinates~\cite{Susskind:1967rg}. Here, we use Susskind's method to transform our calculated wave function from the rest frame into the light-front.

In the light-front frame, a heavy meson is moving with an infinite four-momentum $(E,P)$ in the $z$-direction, whereas the four-momentum for a valence quark inside such a heavy meson is $(k_0,\emph{k})$. Therefore, the quark's four-momentum in the rest frame $(p_0,\emph{p})$ is given by the inverse Lorentz boost $p=L(p \leftarrow k)k$, or equivalently 
\begin{align}
\label{eq:Boost1}
p_{0} &=\frac{E}{M}k_{0}-\frac{P}{M} k_{z} \,, \nonumber \\
p_{z} &=\frac{E}{M} k_{z}-\frac{P}{M} k_{0} \,, \nonumber \\
p_{\perp} &=k_{\perp} \,.
\end{align}

According to Ref.~\cite{Dziembowski:1994jq}, the longitudinal momentum of a quark in the light-front, $k_z$, can be expressed in terms of a fraction $\zeta$,
\begin{equation}
k_{z}=\zeta P \quad\text{with}\quad \sum \zeta=1 \,.
\end{equation}
Then, since $P\rightarrow \infty$ and $\zeta$ should be positive, the quark's on-shell energy can be expanded as
\begin{equation}
k_{0} = \sqrt{\zeta^{2} P^{2}+k_{\perp}^{2}+m_q^{2}} \simeq\zeta P+\frac{k_{\perp}^{2}+m_q^{2}}{2 \zeta P} \,.
\end{equation}
Similarly, the meson's energy can be written as
\begin{equation}
E=\sqrt{P^{2}+M^{2}}\simeq P+\frac{M^{2}}{2 P} \,.
\end{equation}

The inverse Lorentz boost of Eq.~(\ref{eq:Boost1}) can then be rewritten as
\begin{align}
\label{eq:Boost2}
p_{0} &=\frac{1}{2}\left(\zeta M+\frac{p_{\perp}^{2}+m_q^{2}}{\zeta M}\right) \,, \nonumber \\
p_{z} &=\frac{1}{2}\left(\zeta M-\frac{p_{\perp}^{2}+m_q^{2}}{\zeta M}\right) \,, \nonumber \\
p_{\perp} &=k_{\perp} \,.
\end{align}

Noticing that, in the light-front limit, the longitudinal fraction $\zeta=k_z/P$ can be replaced with the light-front fraction $x=k^+/P^+$, \emph{i.e.},
\begin{align}
\label{eq:Boost3}
p_{0} &=\frac{1}{2}\left(p^{+}+p^{-}\right)=\frac{1}{2}\left(x M+\frac{p_{\perp}^{2}+m_q^{2}}{x M}\right) \,, \nonumber \\
p_{z} &=\frac{1}{2}\left(p^{+}-p^{-}\right)=\frac{1}{2}\left(x M-\frac{p_{\perp}^{2}+m_q^{2}}{x M}\right) \,, \nonumber \\
p_{\perp} &=k_{\perp} \,.
\end{align}

In the light-front limit, the spin and orbital angular momentum component of the LFWF, denoted as $\mathcal{R}_{\lambda_q \lambda_{\bar{q}}}^{J J_z}$, is derived using the interaction-independent Melosh transformation. This transformation is applied in the instant form, resulting in the following expressions ~\cite{Arifi:2024mff}, namely:
\begin{equation}
\mathcal{R}_{\lambda_q \lambda_{\bar{q}}}^{00}=\frac{1}{\sqrt{2} \sqrt{\mathcal{A}^2+\boldsymbol{k}_{\perp}^2}}\left(\begin{array}{cc}
k^L & \mathcal{A} \\
-\mathcal{A} & -k^R
\end{array}\right),
\end{equation}
for pseudoscalar mesons, and
\begin{equation}
\begin{aligned}
& \mathcal{R}_{\lambda_q \lambda_{\bar{q}}}^{11}=\frac{1}{\sqrt{\mathcal{A}^2+\boldsymbol{k}_{\perp}^2}}\left(\begin{array}{cc}
\mathcal{A}+\frac{k_{\perp}^2}{\mathcal{D}_0} & k^R \frac{\mathcal{M}_1}{\mathcal{D}_0} \\
-k^R \frac{\mathcal{M}_2}{\mathcal{D}_0} & -\frac{\left(k^R\right)^2}{\mathcal{D}_0}
\end{array}\right), \\
& \mathcal{R}_{\lambda_q \lambda_{\bar{q}}}^{10}=\frac{1}{\sqrt{2} \sqrt{\mathcal{A}^2+\boldsymbol{k}_{\perp}^2}}\left(\begin{array}{cc}
k^L \frac{\mathcal{M}}{\mathcal{D}_0} & \mathcal{A}+\frac{2 k_{\perp}^2}{\mathcal{D}_0} \\
\mathcal{A}+\frac{2 k_{\perp}^2}{\mathcal{D}_0} & -k^R \frac{\mathcal{M}}{\mathcal{D}_0}
\end{array}\right), \\
& \mathcal{R}_{\lambda_q \lambda_{\bar{q}}}^{1-1}=\frac{1}{\sqrt{\mathcal{A}^2+\boldsymbol{k}_{\perp}^2}}\left(\begin{array}{cc}
-\frac{\left(k^L\right)^2}{\mathcal{D}_0} & k^L \frac{\mathcal{M}_2}{\mathcal{D}_0} \\
-k^L \frac{\mathcal{M}_1}{\mathcal{D}_0} & \mathcal{A}+\frac{k_{\perp}^2}{\mathcal{D}_0}
\end{array}\right),
\end{aligned}
\end{equation}
for vector mesons; where $k^{R(L)}=k_x \pm i k_y$, $\mathcal{A}=(1-x) m_1+x m_2$, $D_0=M+m_q+m_{\bar{q}}$, $\mathcal{M}_1=xM+m_1$, $\mathcal{M}_2=(1-x)M+$ $m_2$, and $\mathcal{M}=\mathcal{M}_2-\mathcal{M}_1$.
$\mathcal{R}_{\lambda_q \lambda_{\bar{q}}}^{J J_z}$ is normalized as
\begin{equation}
\sum_{\lambda_q, \lambda_{\bar{q}}} \mathcal{R}_{\lambda_q \lambda_{\bar{q}}}^{J J_z \dagger} \mathcal{R}_{\lambda_q \lambda_{\bar{q}}}^{J J_z}=1 \,.
\end{equation}

It should be noted that, according to Refs.~\cite{Wu:2022iiu, jafar2025}, the meson mass $M$ should be replaced with the so-called invariant mass $M_0$, which is defined as
\begin{equation}
M_0^2=\frac{\boldsymbol{k}_{\perp}^2+m_q^2}{x}+\frac{\boldsymbol{k}_{\perp}^2+m_{\bar{q}}^2}{1-x} \,.
\end{equation}
Then, we obtain the meson's LFWF as follows
\begin{equation}\label{lightcone}
\Psi^{LC}_{\lambda_q \lambda_{\bar{q}}, JJ_z}(x,\vec{p}_{\perp})= \Phi^{LC}_{JJ_z}(x,\vec{p}_{\perp})\mathcal{R}_{\lambda_q \lambda_{\bar{q}}}^{J J_z} \,,
\end{equation}
where
\begin{equation}\label{lightcone2}
\Phi^{LC}_{JJ_z}(x,\vec{p}_{\perp})=\varphi_{JJ_z}(\vec{p}\,)\sqrt{\frac{\partial p_z}{\partial x}} \,.
\end{equation}
$\varphi_{JJ_z}(\vec{p}\,)$ is the meson momentum wave function calculated with GEM and $\sqrt{\partial p_z/\partial x}$ results from the coordinate transformation for $p_z$,
\begin{equation}
\frac{\partial p_z}{\partial x}=\frac{M_0}{4 x(1-x)}\left[1-\frac{\left(m_q^2-m_{\bar{q}}^2\right)^2}{M_0^4}\right] {\,.}
\end{equation}

With the above results, the meson's valence quark PDF is obtained by integrating out the transverse momentum degrees-of-freedom as
\begin{equation}
\begin{aligned}
u_q(x,\zeta_H)&=\sum_{\lambda_q, \lambda_{\bar{q}}}\int d^{2} \vec{p}_{\perp}\left|\Psi^{LC}_{\lambda_q \lambda_{\bar{q}}, JJ_z}(x,\vec{p}_{\perp})\right|^2 \\
&=\int d^{2} \vec{p}_{\perp}\left|\Phi^{LC}_{JJ_z}(x,\vec{p}_{\perp})\right|^{2} \,,
\end{aligned}
\end{equation}
where $\zeta_H$ is defined as the hadron scale. The final PDFs satisfy the following conditions
\begin{subequations}
\begin{align}
& \int_0^1 dx\, u_q(x,\zeta_H)=1 \,, \\
& \int_0^1 dx\, x[(u_q(x,\zeta_H)+u_{\bar{q}}(x,\zeta_H)]=1 \,.
\end{align}
\end{subequations}

%%%%%%%%%%%%%%%%%%%%%%%%%%%%%%%%%%%%%%%%%%%%%%%%%%%%%%%%%%%%%%%%%%%%%%%%%%%%%%%%%

\section{RESULTS}
\label{sec:Results}

\begin{figure}[!t]
\centering
\includegraphics[width=0.45\textwidth]{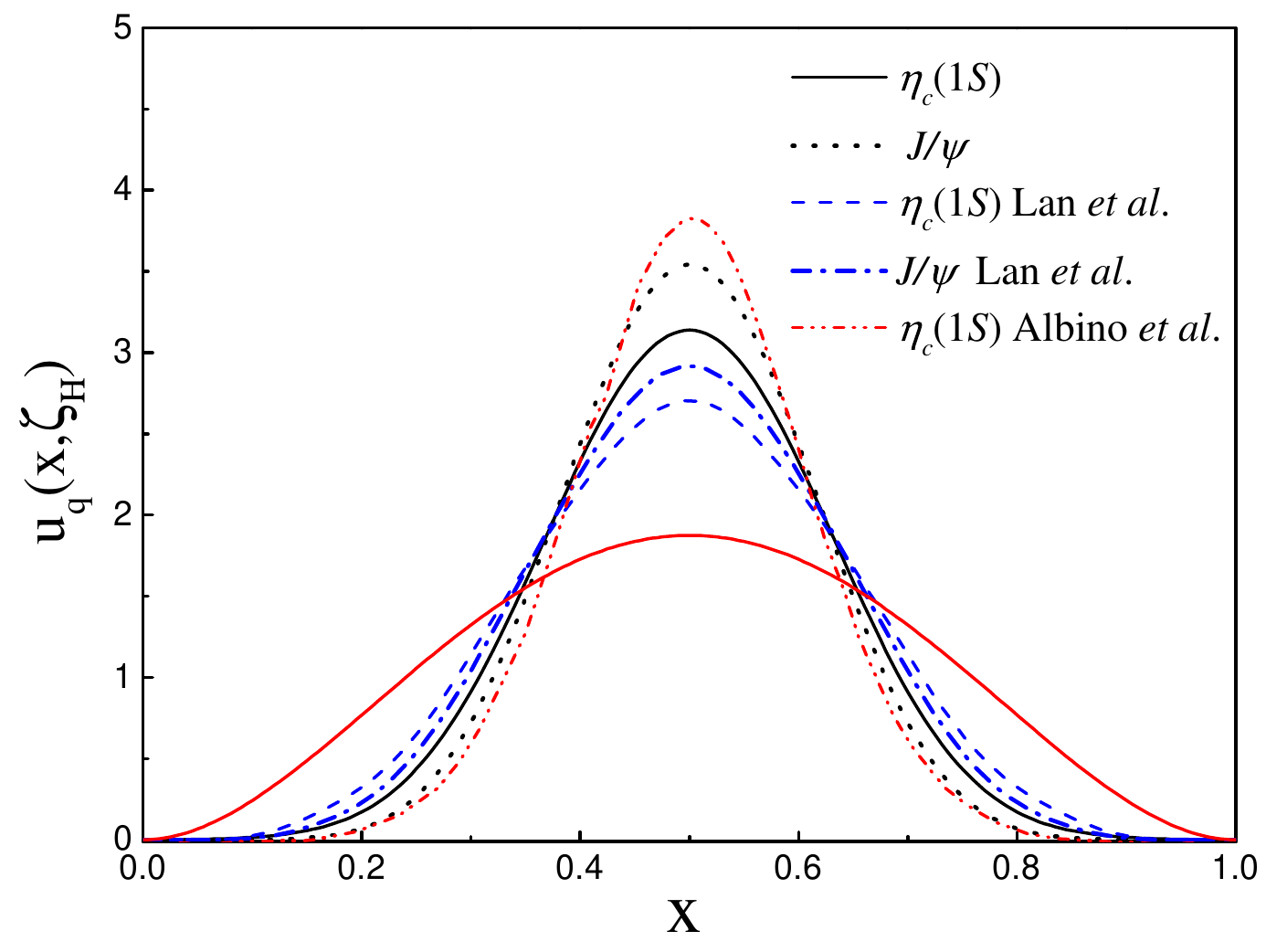}
\caption{\label{fig:ccPDF} Valence $c$ quark PDFs at the hadron scale, $\zeta_H$, of $\eta_c(1S)$ (black solid) and $J/\psi$ (black dotted) mesons. The blue dashed and dash-dotted lines correspond, respectively, to the PDFs of $\eta_c(1S)$ and $J/\psi$ given in Ref.~\cite{Lan:2019img}. The red double-dot-dashed line is provided by Ref.~\cite{Albino:2022gzs}. For comparison, the scale-free parton-like profile: $q(x)=30x^2(1-x)^2$, is also shown as a red solid line.}
\end{figure}

The valence quark PDFs of pseudoscalar and vector ground state charmonia are depicted in Fig.~\ref{fig:ccPDF}. The (black) solid line corresponds to the valence $c$ quark PDF of the $\eta_c(1S)$ meson and the (black) dotted line refers to the $J/\psi$ case. As indicated in the figure, the remaining curves belong to other phenomenological calculations: Lan \emph{et al.} uses a basis light-front quantization (BLFQ) approach to a QCD-inspired effective Hamiltonian~\cite{Lan:2019img}, whereas Albino \emph{et al.} employ an algebraic model for the quark Dyson-Schwinger equation and for the pseudoscalar meson Bethe-Salpeter equation~\cite{Albino:2022gzs}. From Fig.~\ref{fig:ccPDF}, the following observations are particularly noteworthy: the PDFs of $c$ quark in $\eta_c(1S)$ and $J/\psi$ mesons are narrower than the scale-free parton-like profile, $q(x)=30x^2(1-x)^2$; and as expected, the PDFs are centered with respect to the momentum fraction $x=0.5$; the $J/\psi$'s valence quark PDF is slightly higher at $x=0.5$ and lower at the edges than that of the $\eta_c(1S)$ meson. Compared with recent results reported by other approaches, everything seems to indicate that our curves are in reasonable agreement with existing analyses, and therefore they follow the widely expected patterns. Interestingly, the structural differences between the $\eta_c(1S)$ and $J/\psi$ PDFs seen in our results are mirrored in the work of Lan \emph{et al.}, even though they employ an entirely different method. Note, finally, that the corresponding anti-quark PDF is simply obtained as $\bar{Q}(x;\zeta_H) = q(1-x,\zeta_H)$, because at the hadron scale $\zeta_H$, the dressed valence quarks express all hadron properties; in particular, they fully carry the meson momentum.

\begin{figure}[!t]
\centering
\includegraphics[width=0.45\textwidth]{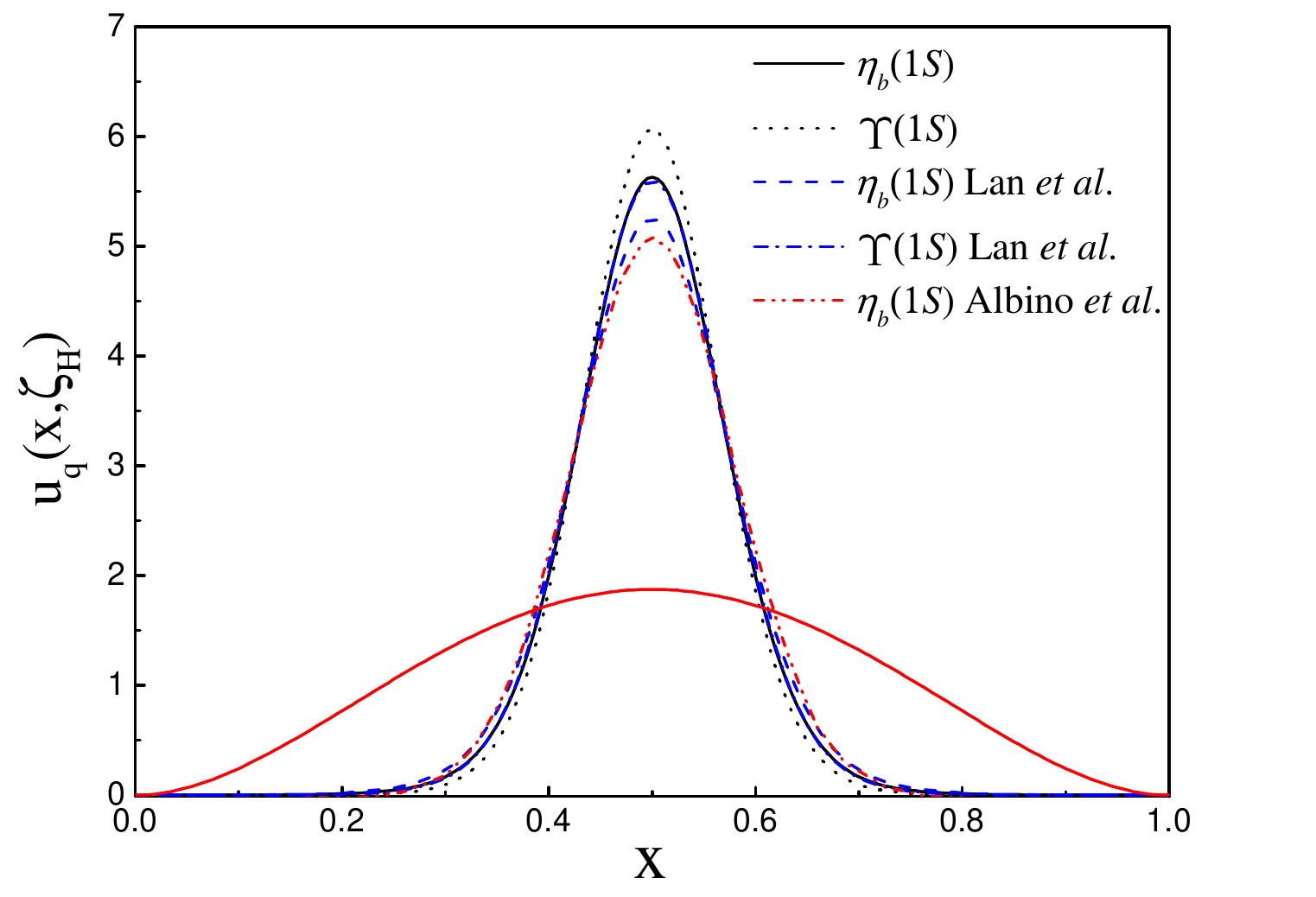}
\caption{\label{fig:bbPDF} Valence $b$ quark PDFs at the hadron scale, $\zeta_H$, of $\eta_b(1S)$ (black solid) and $\Upsilon(1S)$ (black dotted) mesons. The blue dashed and dash-dotted lines correspond, respectively, to the PDFs of $\eta_b(1S)$ and $\Upsilon(1S)$ given in Ref.~\cite{Lan:2019img}. The red double-dot-dashed line is provided by Ref.~\cite{Albino:2022gzs}. For comparison, the scale-free parton-like profile: $q(x)=30x^2(1-x)^2$, is also shown as a red solid line.}
\end{figure}

The valence quark PDFs of pseudoscalar and vector ground state bottomonia ($b{\bar b}$) are depicted in Fig.~\ref{fig:bbPDF}. The (black) solid line corresponds to the valence $b$ quark PDF of $\eta_b(1S)$ meson and the (black) dotted line refers to the $\Upsilon(1S)$ case. As before, the remaining curves belong to other phenomenological calculations~\cite{Lan:2019img, Albino:2022gzs}. The PDFs of $b$ quark in the $\eta_b(1S)$ and $\Upsilon(1S)$ mesons are narrower than those reported for the $\eta_c(1S)$ and $J/\psi(1S)$ mesons, and thus they are even narrower with respect to the scale-independent parton-like distribution, $q(x) = 30x^2(1-x)^2$. The PDFs are again centered around the momentum fraction $x=0.5$. The valence quark PDF of the $\Upsilon(1S)$ meson is slightly higher at $x = 0.5$ and lower at the edges compared to the $\eta_b(1S)$ meson's one. Again, compared to recent results from other methods, our curves appear to align reasonably well with predicted patterns; for example, the differences between the $\eta_b(1S)$ and $\Upsilon(1S)$ PDFs found in our study are similarly reflected in Ref.~\cite{Lan:2019img}. {However, it is very interesting to notice that, while the $\eta_c(1S)$ results in Ref.~\cite{Albino:2022gzs} are narrower than ours and those in Ref. ~\cite{Lan:2019img}, the $\eta_b(1S)$ results are the opposite. This means the framework in Ref.~\cite{Albino:2022gzs} is less sensitive on quark mass than those in Ref. ~\cite{Lan:2019img} and ours. Namely, with $c$ quark changed to $b$, which is almost 3 times heavier, the peak value (at x=0.5) only increases about 30\%, while it's over 80\% for Ref. ~\cite{Lan:2019img} and our model.}
Lastly, note once more that the corresponding anti-quark PDF can be obtained as $\bar{Q}(x;\zeta_H) = q(1-x,\zeta_H)$, since at the hadron scale $\zeta_H$ the dressed valence quarks account for all hadron properties, including the full hadron momentum.

\begin{figure}[!t]
\centering
\includegraphics[width=0.45\textwidth]{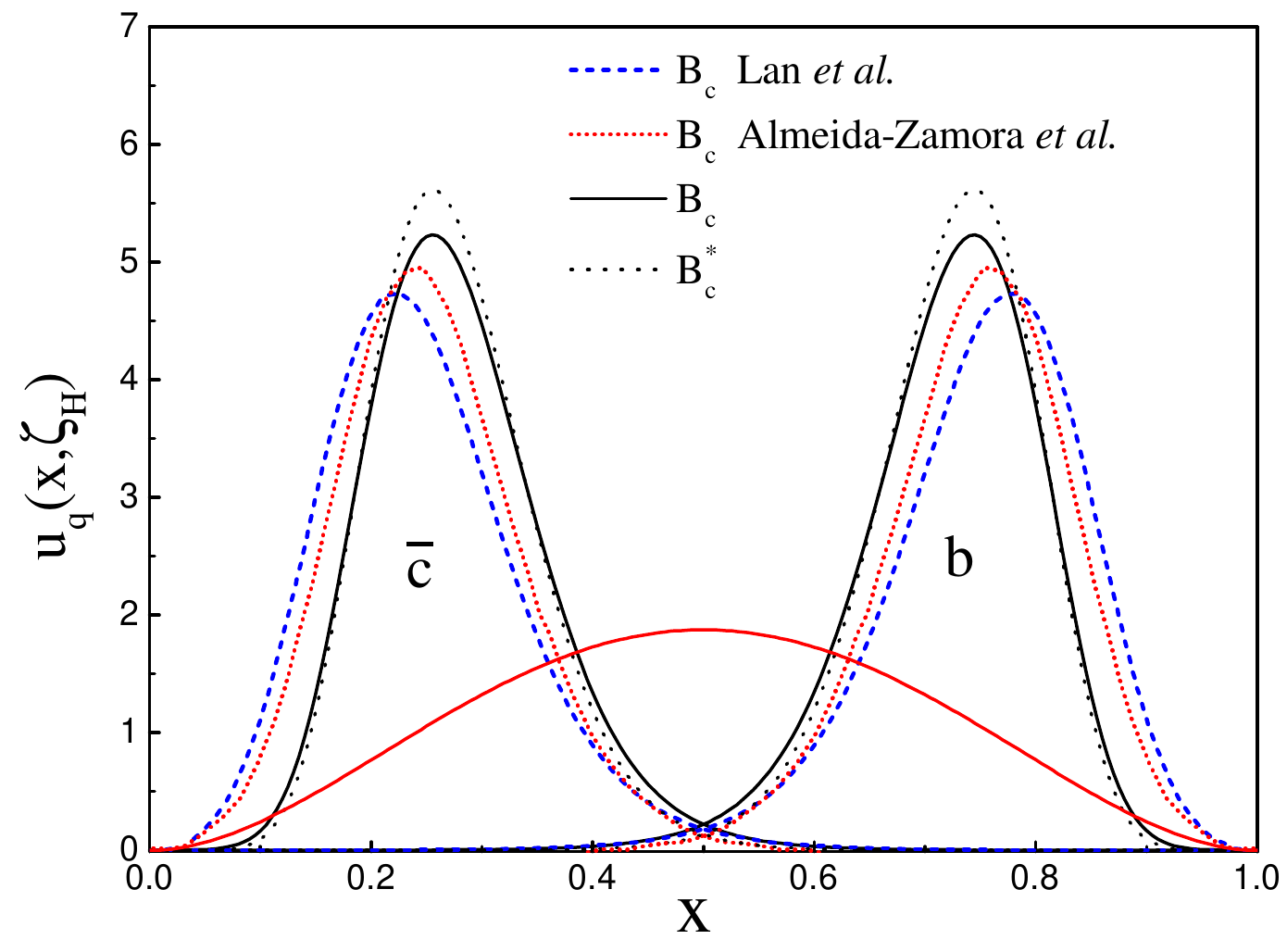}
\caption{\label{fig:BcPDF} Valence $b$ quark and $\bar c$ anti-quark PDFs at the hadron scale, $\zeta_H$, of $B_c$ (black solid line) and $B_c^\ast$ (black dotted line) mesons. The blue dashed line corresponds to the PDFs reported in Ref.~\cite{Lan:2019img}. The red dotted line is provided by Ref.~\cite{Almeida-Zamora:2023bqb}. For comparison, the scale-free parton-like profile: $q(x)=30x^2(1-x)^2$, is also shown as a red solid line.}
\end{figure}

Fig.~\ref{fig:BcPDF} shows the valence $b$ quark and $\bar c$ anti-quark PDFs at the hadron scale, $\zeta_H$, of $B_c$ (black solid line) and $B_c^\ast$ (black dotted line) mesons~\footnote{Note that our model can't distinguish $B_c^-(\bar cb)$ and $B_c^+(c\bar b)$}. The blue dashed line corresponds to the PDFs reported in Ref.~\cite{Lan:2019img} where a BLFQ approach is used. The red dotted line is provided by Ref.~\cite{Almeida-Zamora:2023bqb} in this case, where Albino's algebraic model is used to describe PDFs of heavy-light mesons. As we see, it is obvious that the PDFs of $b$ quark and $\bar c$ antiquark satisfy $\bar{c}(x;\zeta_H)=b(1-x;\zeta_H)$, both in the $B_c$ and $B_c^*$ cases. One can also see that all drawn PDFs are again more pronounced than the conformal parton-like PDF and, obviously, asymmetric with respect to it as they represent heavy-light mesons. As expected, the heavier quark carries larger values of fractional longitudinal momentum than the lighter one. When we compare our results with other theoretical approaches, they are in reasonable agreement. Lan \emph{et al.} predict curves slightly displaced towards the edges of $x$-variable but their point-wise behavior is pretty similar to the ones predicted by us. Ref.~\cite{Almeida-Zamora:2023bqb} predicts curves that are in the middle of ours and those of Lan \emph{et al.}; however, everything seems to indicate that all predictions are in reasonable agreement and, therefore, follow the expected patterns. Especially, to our knowledge currently there is no other valence quark PDF predictions for $B_c^\ast$, ours stands alone.

To delve deeper into the properties of the LFWFs, and the PDFs derived from them, here we proceed by calculating the Mellin moments of these distribution functions. To be specific, the Mellin moments of the LFWFs are defined as
\begin{equation}
\begin{aligned}
\left\langle x^m\right\rangle_{p^2_\perp} &=
\sum_{\lambda_q, \lambda_{\bar{q}}} \int_0^1 dx\, x^m \left|\Psi^{LC}_{\lambda_q \lambda_{\bar{q}}, JJ_z}(x,\vec{p}_{\perp})\right|^2 \\
&
=\int_0^1 dx\, x^m
\left|\Phi^{LC}_{JJ_z}(x,\vec{p}_{\perp})\right|^{2},
\end{aligned}
\end{equation}
we then integrate over $\vec{p}_{\perp}$,
\begin{equation}
\int d^2\vec{p}_{\perp} \left\langle x^m\right\rangle_{p^2_\perp}=\langle x^m\rangle \,,
\end{equation}
in order to compute the corresponding Mellin moments of the PDFs.

\begin{table}[t]
\centering
\caption{\label{tab:Mellin-LCWFs} Calculated Mellin moments of light-front wave functions at the hadron scale, $\langle x^m \rangle_{p_\perp^2}$, with $m=0,1,2,3,4$, and $p_{\perp}^2=0.0,0.1, 0.2\,\mathrm{GeV}^2$. All quantities are given in $\mathrm{GeV}^{-2}$.}
\begin{tabular}{llccccc}
\hline
\hline
\noalign{\vskip 0.1true cm}
$\langle x^m\rangle_{p^2_\perp}\;\;\;$ & $\;\;p_{\perp}^2$&$\;\;\;\;m=0$ & $\;\;\;\;1\;\;\;\;$ & $\;\;\;\;2\;\;\;\;$ & $\;\;\;\;3\;\;\;\;$ & $\;\;\;\;4\;\;\;\;$  \\
\noalign{\vskip 0.1true cm}
\hline
\noalign{\vskip 0.05true cm}
$\eta_c(1S)$ & $0.0$ & 0.738 & 0.369 & 0.195 & 0.108 & 0.062   \\
         & $0.1$ & 0.561 & 0.281 & 0.148 & 0.082 & 0.048   \\
         & $0.2$ & 0.431 & 0.216 & 0.114 & 0.063 & 0.037   \\
           \noalign{\vskip 0.05true cm}
   \hline
   \noalign{\vskip 0.05true cm}
$J/\psi$ & $0.0$ & 0.947 & 0.474 & 0.248 & 0.135 & 0.076   \\
         & $0.1$ & 0.682 & 0.341 & 0.178 & 0.097 & 0.055   \\
         & $0.2$ & 0.497 & 0.249 & 0.130 & 0.071 & 0.040  \\
     \noalign{\vskip 0.05true cm}
   \hline
   \noalign{\vskip 0.05true cm}
$\eta_b(1S)$ & $0.0$ & 0.310 & 0.155 & 0.079 & 0.041 & 0.022   \\
         & $0.1$ & 0.276 & 0.138 & 0.070 & 0.036 & 0.019   \\
         & $0.2$ & 0.246 & 0.123 & 0.062 & 0.033 & 0.017   \\
        \noalign{\vskip 0.05true cm}
   \hline
   \noalign{\vskip 0.05true cm}
$\Upsilon(1S)$ & $0.0$ & 0.351 & 0.179 & 0.091 & 0.047 & 0.025   \\
           & $0.1$ & 0.314 & 0.157 & 0.080 & 0.041 & 0.022   \\
           & $0.2$ & 0.277 & 0.138 & 0.070 & 0.036 & 0.019   \\
        \noalign{\vskip 0.05true cm}
   \hline
   \noalign{\vskip 0.05true cm}
$B_c(\bar{c})$   & $0.0$ & 0.576 & 0.151 & 0.043 & 0.013 & 0.004   \\
                 & $0.1$ & 0.467 & 0.124 & 0.025 & 0.011 & 0.004   \\
                 & $0.2$ & 0.379 & 0.102 & 0.039 & 0.009 & 0.003   \\
         \noalign{\vskip 0.05true cm}
   \hline
   \noalign{\vskip 0.05true cm}
$B^*_c(\bar{c})$ & $0.0$ & 0.660 & 0.173 & 0.048 & 0.014 & 0.005   \\
                 & $0.1$ & 0.523 & 0.138 & 0.039 & 0.012 & 0.004   \\
                 & $0.2$ & 0.417 & 0.111 & 0.032 & 0.010 & 0.003   \\
          \noalign{\vskip 0.05true cm}
   \hline
     \noalign{\vskip 0.05true cm}
$B_c(b)$         & $0.0$ & 0.576 & 0.425 & 0.317 & 0.238 & 0.180    \\
                 & $0.1$ & 0.467 & 0.343 & 0.255 & 0.191 & 0.144    \\
                 & $0.2$ & 0.379 & 0.278 & 0.206 & 0.154 & 0.116    \\
         \noalign{\vskip 0.05true cm}
   \hline
   \noalign{\vskip 0.05true cm}
$B^*_c(b)$       & $0.0$ & 0.660 & 0.488 & 0.363 & 0.272 & 0.206   \\
                 & $0.1$ & 0.523 & 0.355 & 0.286 & 0.214 & 0.161   \\
                 & $0.2$ & 0.417 & 0.306 & 0.226 & 0.169 & 0.127   \\
          \noalign{\vskip 0.05true cm}
\hline
\hline
\end{tabular}
\end{table}

\begin{table}[!t]
\centering
\caption{\label{tab:Mellin-PDFs} Calculated Mellin moments of parton distribution functions at the hadron scale, $\langle x^m \rangle$, with $m=0,1,2,3,4$.}
\begin{tabular}{lccccc}
\hline
\hline
\noalign{\vskip 0.1true cm}
  &$\;\;\;\;\langle x^0\rangle\;\;\;\;$  & $\;\;\;\;\langle x^1\rangle\;\;\;\;$ & $\;\;\;\;\langle x^2\rangle\;\;\;\;$ & $\;\;\;\;\langle x^3\rangle\;\;\;\;$ & $\;\;\;\;\langle x^4\rangle\;\;\;\;$  \\
\noalign{\vskip 0.1true cm}
\hline
\noalign{\vskip 0.05true cm}
  $\eta_c(1S)$       & 1.000 &0.500 &0.266    &0.148 & 0.087   \\
  $J/\psi$       & 1.000 &0.500 &0.262    &0.143 & 0.081    \\
  $\eta_b(1S)$       & 1.000 &0.500 & 0.256   &0.133 & 0.071    \\
  $\Upsilon(1S)$     & 1.000 &0.500 & 0.255   &0.132 & 0.070    \\
  $B_c(\bar{c})$ & 1.000 &0.278 & 0.084   &0.028 & 0.010    \\
$B^*_c(\bar{c})$ & 1.000 &0.275 & 0.081   &0.025 & 0.009    \\
$B_c(b)$         & 1.000 &0.722 & 0.527   &0.390 & 0.291   \\
$B^*_c(b)$       & 1.000 &0.725 & 0.531   &0.392 & 0.293   \\
     \noalign{\vskip 0.1 true cm}
\hline
\hline
\end{tabular}
\end{table}

\begin{figure}[!t]
\centering
\includegraphics[width=0.45\textwidth]{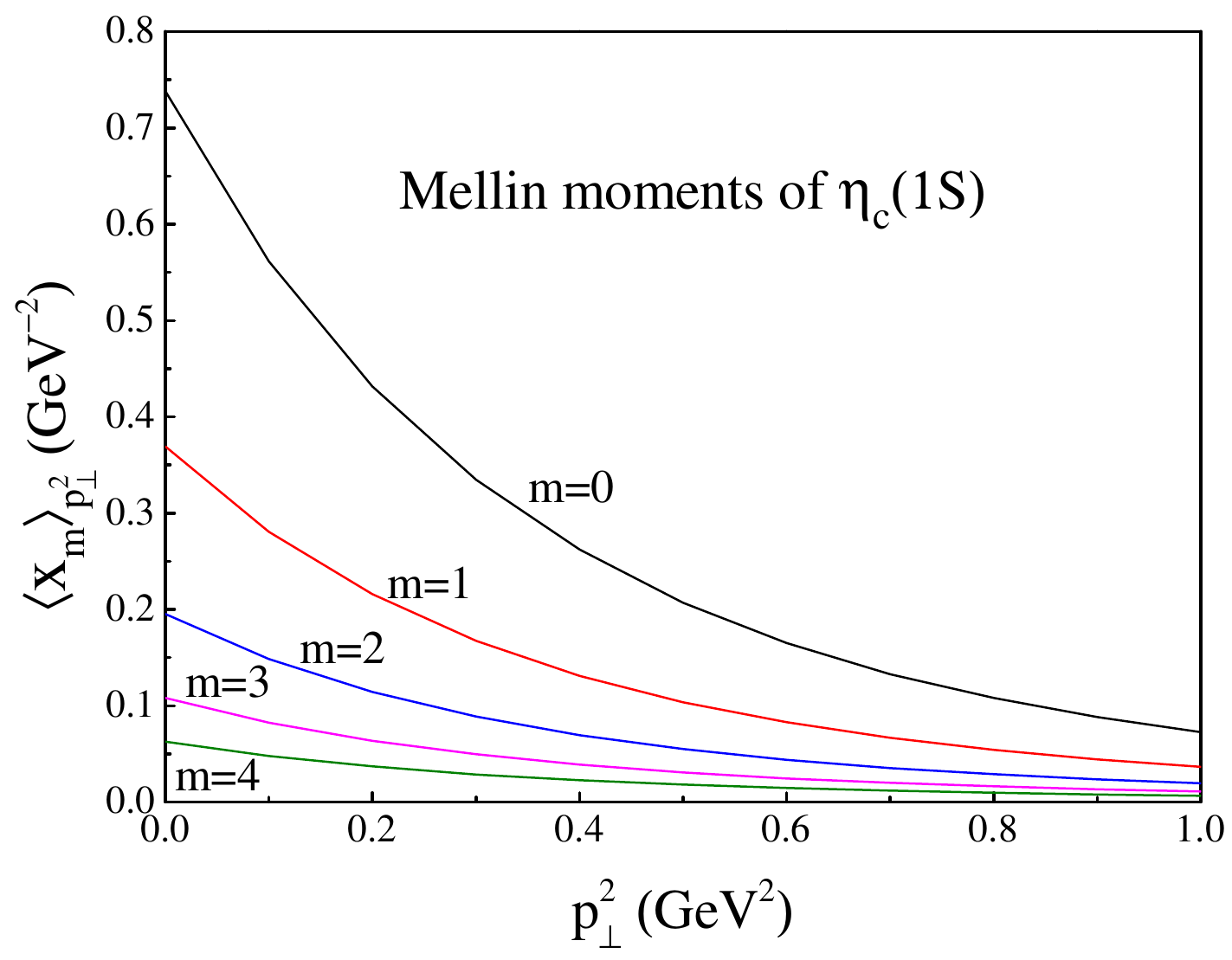}
\caption{\label{etacmm} Mellin moments of $\eta_c(1S)$ light-front wave functions at the hadron scale.}
\end{figure}

\begin{figure}[!t]
\centering
\includegraphics[width=0.45\textwidth]{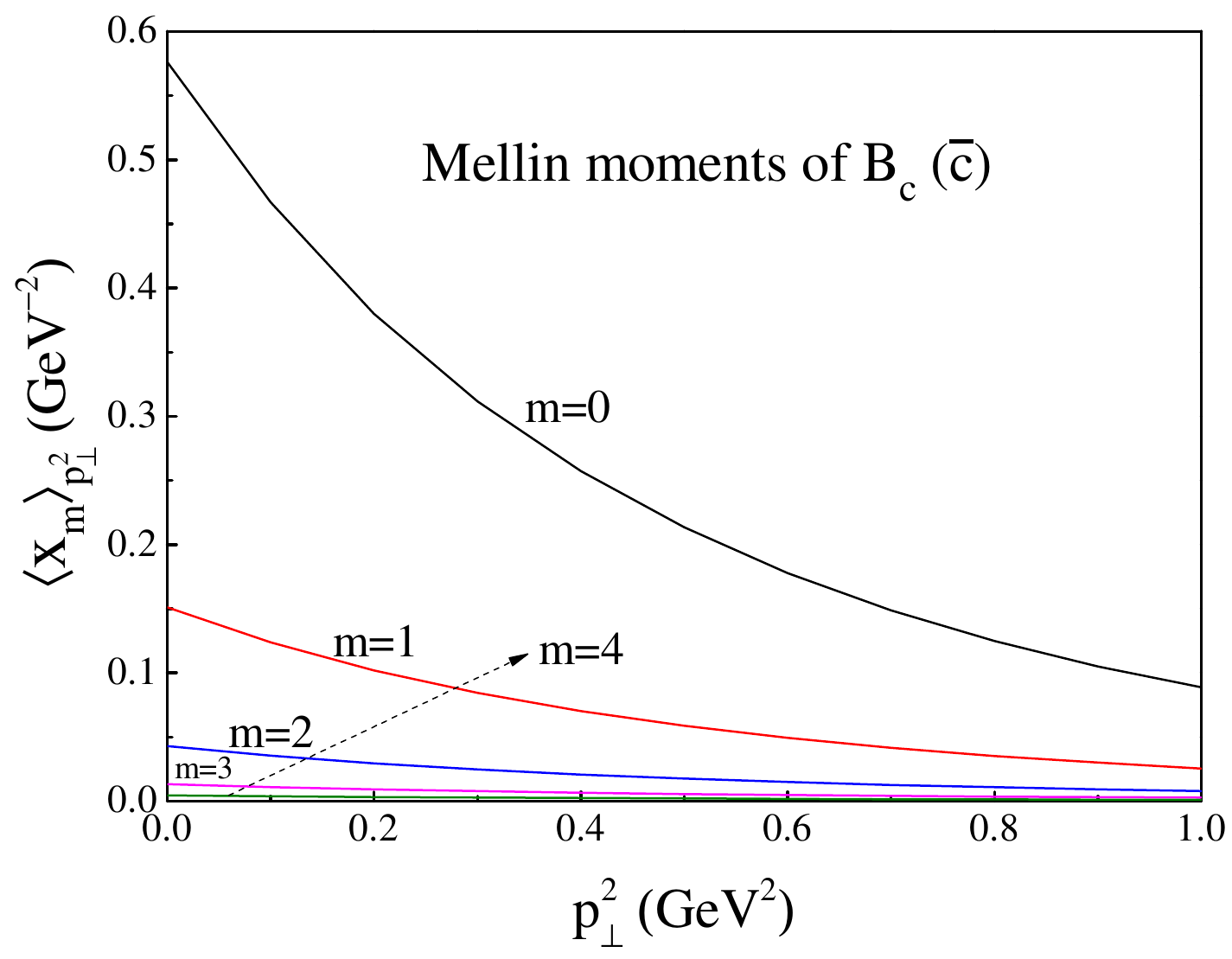}
\caption{\label{bccmm} Mellin moments of $B_c(\bar{c})$ light-front wave functions at the hadron scale.}
\end{figure}

\begin{figure}[!t]
\centering
\includegraphics[width=0.45\textwidth]{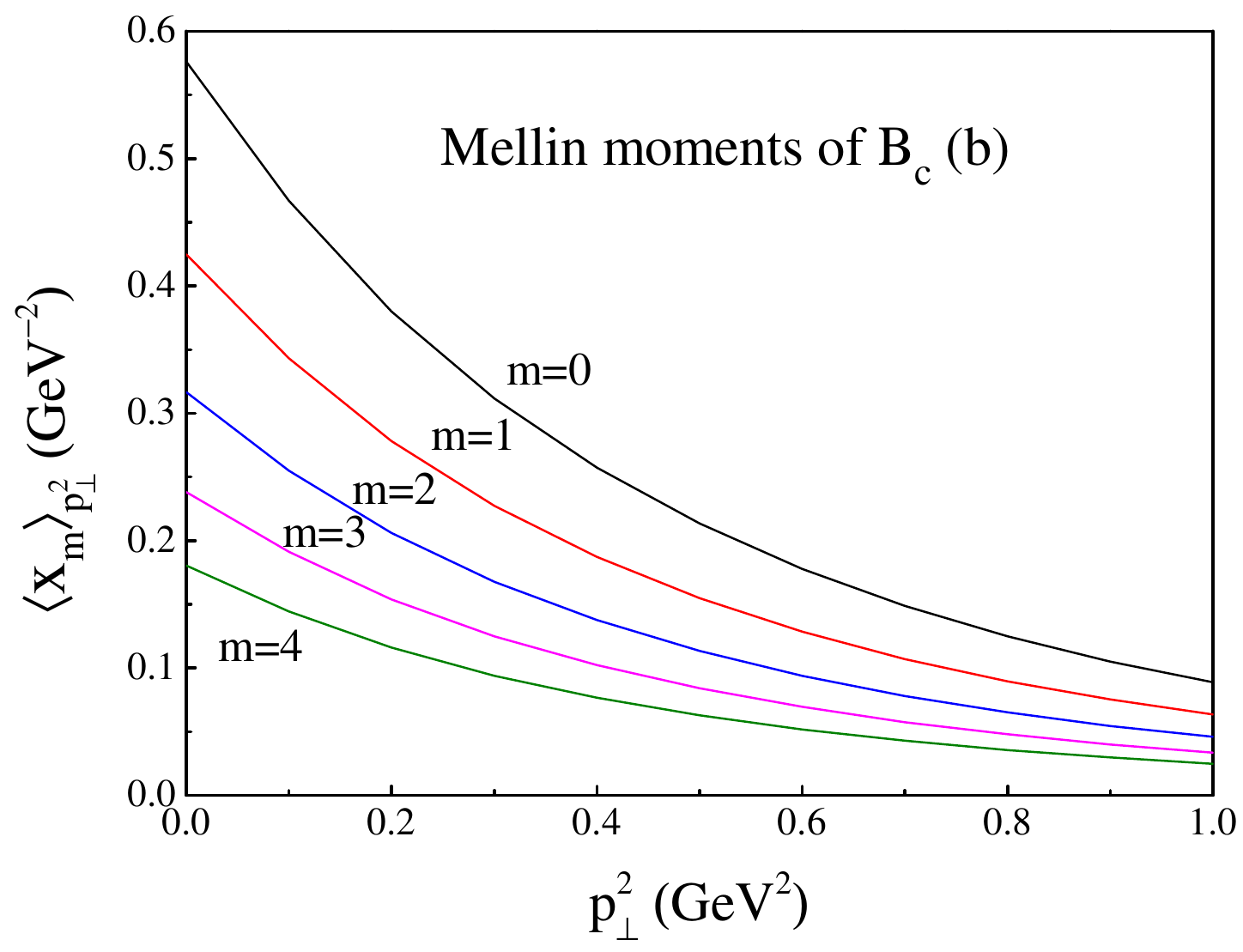}
\caption{\label{bcbmm} Mellin moments of $B_c(\bar{b})$ light-front wave functions at the hadron scale.}
\end{figure}

Table~\ref{tab:Mellin-LCWFs} shows the calculated Mellin moments of LFWFs at the hadron scale, $\langle x^m \rangle_{p_\perp^2}$, with $m=0,1,2,3,4$, and $p_{\perp}^2=0.0,0.1, 0.2\,\mathrm{GeV}^2$. As one can find easily, they systematically fall-off towards zero, while the former is always larger than the latter. Moreover, for any meson, the value of a given moment decreases as $p_\perp^2$ increases. Furthermore, the moments for $\eta_b(1S)$ and $\Upsilon(1S)$ mesons exhibit a smoother dependence on $p_\perp^2$, with the moments at $p_\perp^2=0.2\,\text{GeV}^2$ being nearly $80\%$ of the values at $p_\perp^2=0.0\,\text{GeV}^2$. However, for $\eta_c(1S)$ and $J/\psi$ mesons, this percentage is approximately $55\%$, and is around $60\%$ for $B_c$ and $B_c^\ast$ mesons. {To provide a clearer visual representation, we plot the representative trends in Figs.~\ref{etacmm}, ~\ref{bccmm} and ~\ref{bcbmm}.}

\begin{figure}[!t]
\centering
\includegraphics[width=0.45\textwidth]{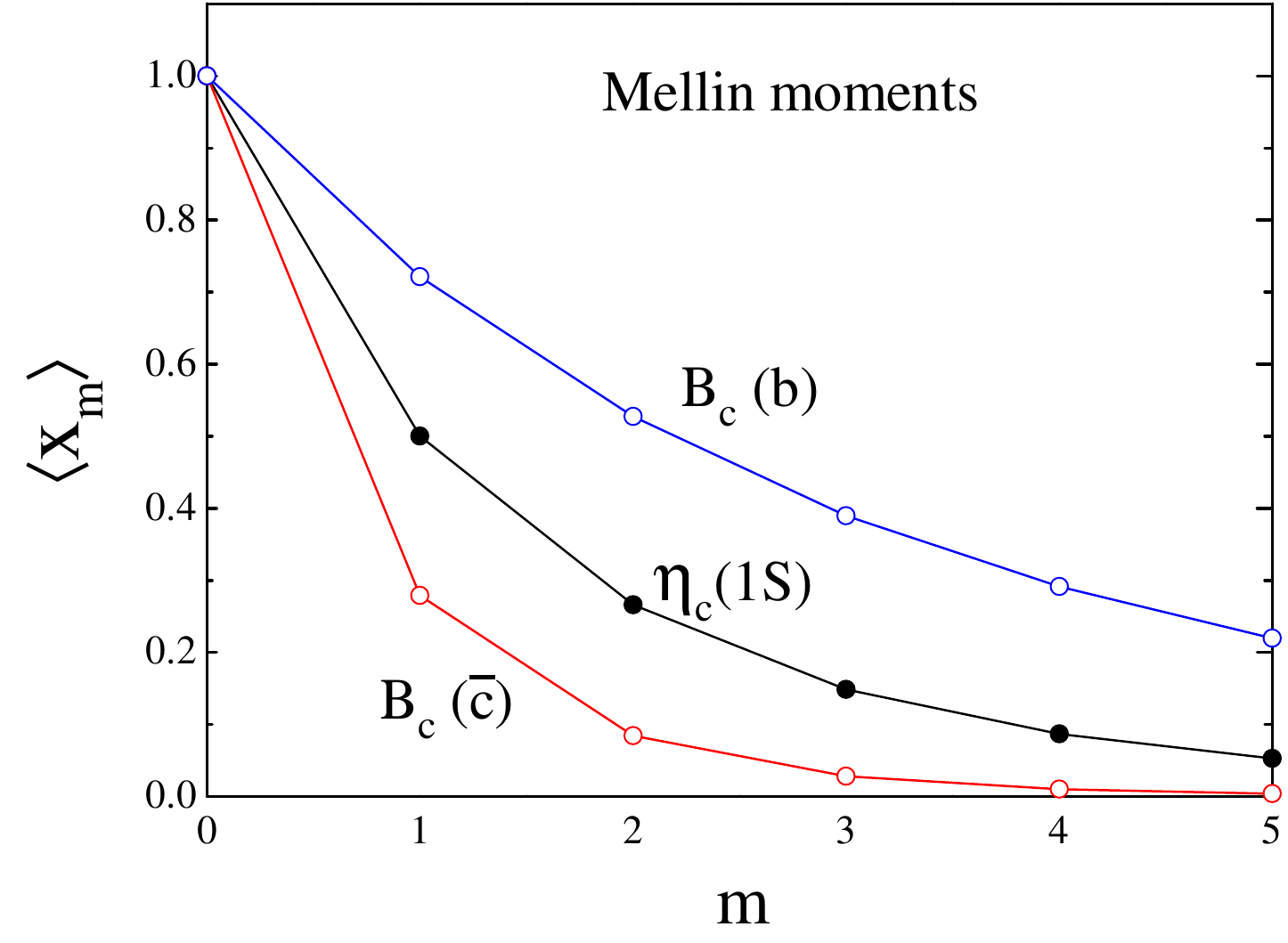}
\caption{\label{tableiv} Mellin moments of selected meson parton distribution functions at the hadron scale.}
\end{figure}

Finally, we show in Table~\ref{tab:Mellin-PDFs} the Mellin moments of associated PDFs at the hadron scale, $\langle x^m \rangle$, with $m=0,1,2,3,4$. Again, as one can notice, they systematically fall-off, being always the former larger than the latter. For hidden-flavor heavy quarkonia, they follow the expected trend observed already in lighter ground state pseudoscalar mesons, such as the pion, reflecting symmetric point-wise behavior with respect to $x$. The PDFs of heavy-light mesons are asymmetric with respect to $x$ and thus this is exhibited in the numerical values of their Mellin moments. {Once again, a visual representation is plotted in Fig.~\ref{tableiv}.}

%%%%%%%%%%%%%%%%%%%%%%%%%%%%%%%%%%%%%%%%%%%%%%%%%%%%%%%%%%%%%%%%%%%%%%%%%%%%%%%%%
%%%%%%%%%%%%%%%%%%%%%%%%%%%%%%%%%%%%%%%%%%%%%%%%%%%%%%%%%%%%%%%%%%%%%%%%%%%%%%%%%

\section{SUMMARY}
\label{sec:Summary}

In this study, we used a constituent quark model (CQM) and Susskind's method to calculate the parton distribution functions (PDFs) of mesons composed of $c$ or $b$ quarks, focusing on all the pseudoscalar and vector ground states. The CQM used herein has been revealed to be a reliable tool for modeling heavy hadrons, describing a wide range of physical observables such as spectra, together with strong, weak and electromagnetic decays and reactions. So, the advantage is that all parameters in this work have already been determined in previous studies, we do not do fine tuning or introduce new ones. Our approach combines the dynamics of one-gluon exchange and confinement, enabling us to solve the Schr\"odinger equation by using a Rayleigh-Ritz variational method, whose solution is assumed to be of Gaussian form, to determine the eigenenergies and eigenfunctions of mesons. By transforming these wave functions into the light-front frame, we provide a clear representation of how the meson's four-momentum is distributed among its quarks.

Our results confirm the robustness of the CQM for predicting PDFs of heavy mesons, and reveal key insights into the structure information of heavy quarkonia, namely, $\eta_c(1S)$,  $J/\psi$, $\eta_b(1S)$, and $\Upsilon(1S)$. The PDFs are shown to be narrower compared to the scale-free parton-like distribution, with the quarks carrying most of the meson's momentum centered at $x = 0.5$. Pseudoscalar meson PDFs are broader than vector ones at the edges, making the highest value at $x=0.5$ for vectors larger than that of pseudoscalars. The structural differences between the PDFs of quarkonia align with those reported by other theoretical approaches, despite differences in methodology, confirming that this CQM is a suitable framework for studying heavy meson properties via PDFs.

Furthermore, we extend our analysis to heavy-light systems, such as the $B_c$ and $B_c^\ast$ mesons. Our findings suggest that the heavier quark tends to carry more of hadron's momentum than its lighter counterpart. All drawn PDFs are more pronounced than the conformal parton-like PDF and, obviously, asymmetric with respect to it since they represent heavy-light mesons. Compared with other theoretical approaches, our results appear to align reasonably well with the expected patterns; among them, as far as we know our predictions for $B_c^*$ are the first.

Last but not least, we also compute some lower Mellin moments of the related light-front wave functions (LFWFs) and PDFs, at the hadron scale. In both cases, they systematically fall-off towards zero, while the former being always larger than the latter. {Moreover, our results show that concerning those of LFWFs, the value of a given moment decreases as $p_\perp^2$ increases.}

%%%%%%%%%%%%%%%%%%%%%%%%%%%%%%%%%%%%%%%%%%%%%%%%%%%%%%%%%%%%%%%%%%%%%%%%%%%%%%%%%

\begin{acknowledgments}

We thank Dr. Ahmad Jafar Arifi for very helpful discussions concerning the Melosh transformations.
Work supported by: Natural Science Foundation of Jiangsu Province (grant no. BK20220122); National Natural Science Foundation of China (grant no. 12233002); China Postdoctoral Science Foundation (grant no. 2024M751369) and Jiangsu Funding Program for Excellent Postdoctoral Talent.
One of the authors (J.S.) is financed by
Ministerio Espa\~nol de Ciencia e Innovaci\'on (grant no. PID2022-140440NB-C22);
Junta de Andaluc\'ia (contract nos. PAIDI FQM-370 and PCI+D+i) under the title: ``Tecnolog\'\i as avanzadas para la exploraci\'on del universo y sus componentes" (code AST22-0001).
\end{acknowledgments}

%%%%%%%%%%%%%%%%%%%%%%%%%%%%%%%%%%%%%%%%%%%%%%%%%%%%%%%%%%%%%%%%%%%%%%%%%%%%%%%%

%\bibliography{HQ_PDF}

%

\end{document}